\documentclass[11pt,a4paper,usenames]{article}  
\usepackage{epsfig,amssymb,amsmath,latexsym} 
\usepackage{jcappub} 

\usepackage{pifont} 
\usepackage{hyperref} 
\usepackage[T1]{fontenc}
 
\makeatletter
\gdef\@fpheader{{\Large CERN-PH-TH-2014-132}}
\makeatother

\usepackage{trajan} 
\newcommand{\be}{\begin{equation}} 
\newcommand{\ee}{\end{equation}} 
\newcommand{\bea}{\begin{eqnarray}} 
\newcommand{\eea}{\end{eqnarray}}

\newcommand{\nb}{\nonumber}

\newcommand{\de}{\partial} 
 
\newcommand\E{{\mathcal{E}}} 
\newcommand\K{{\mathcal{K}}} 
\newcommand\Q{{\cal Q}} 
\renewcommand\S{{\cal S}} 
\newcommand\T{{\cal T}}

\renewcommand\l{\lambda} 

\renewcommand\a{\alpha} 
\newcommand\D{\text{DoF}} 
\renewcommand\b{\beta}

\newcommand\m{\mu} 
\newcommand\n{\nu} 
\newcommand\g{\gamma}

\newcommand\V{{\ensuremath{\cal V}}} 
\newcommand\U{{\ensuremath{\cal U}}} 
 
\newcommand\Hc{{\ensuremath{\cal H}}} 
\newcommand\F{{\ensuremath{\cal F}}} 
\newcommand\tr{\text{Tr}}

\renewcommand\l{\ensuremath{\lambda}}

\newcommand\ba{\begin{array}} 
\newcommand\ea{\end{array}}

\newcommand{\plm}{M_{\text{pl}}^2}

\title{Nonderivative Modified Gravity: a Classification} 
 
\author[a,b]{D. Comelli,}
\author[c,d]{F. Nesti,}
\author[e,f]{L. Pilo} 
\affiliation[a]{INFN - Sezione di Ferrara,  I-35131 Ferrara, Italy}
  \affiliation[b]{CERN, Theory Division, 1211 Geneva, Switzerland} 
\affiliation[c]{Ru\dj er Bo\v skovi\'c Institute, Bijeni\v cka cesta 54, 10000, Zagreb, Croatia}
\affiliation[d]{Gran Sasso Science Institute, viale Crispi 7, I-67100 L'Aquila, Italy}
\affiliation[e]{Dipartimento di Scienze Fisiche e Chimiche,
  Universit\`a di L'Aquila,  I-67010 L'Aquila, Italy}
\affiliation[f]{INFN, Laboratori Nazionali del Gran Sasso, I-67010
  Assergi, Italy} 
\emailAdd{comelli@fe.infn.it}
\emailAdd{fabrizio.nesti@irb.hr}
\emailAdd{luigi.pilo@aquila.infn.it}

\date{\small\today}

\abstract{We analyze the theories of gravity modified by a generic nonderivative potential built
  from the metric, under the minimal requirement of unbroken spatial rotations.  Using the canonical
  analysis, we classify the potentials $V$ according to the number of degrees of freedom (DoF) that
  propagate at the nonperturbative level.  We then compare the nonperturbative results with the
  perturbative DoF propagating around Minkowski and FRW backgrounds.  A generic $V$ implies 6
  propagating DoF at the non-perturbative level, with a ghost on Minkowski background.  There exist
  potentials which propagate 5\,DoF, as already studied in previous works. Here, no $V$ with
  unbroken rotational invariance admitting 4\,DoF is found.  Theories with 3\,DoF turn out to be
  strongly coupled on Minkowski background.  Finally, potentials with only the 2\,DoF of a massive
  graviton exist.  Their effect on cosmology is simply equivalent to a cosmological constant.
  Potentials with 2 or 5 DoF and explicit time dependence appear to be a further viable
  possibility.}

\begin{document} 
 
\maketitle 
 
\section{Introduction}   
In the recent years, there has been a substantial progress in understanding possible modifications
of Einstein General Relativity at large distances.  The quest is on for a theory of gravity which
has a massive graviton in the spectrum at the linearized level, thus realizing a full nonlinear
theory of gravity modified at large distance.

The main goal of the present investigation is to study systematically the theories of massive
gravity obtained by adding a nonderivative potential of the metric components $g_{\m\n}$ to the
Einstein Hilbert (EH) action\footnote{For the rest of the paper we will set $\plm=1$ when not
  important for our intents.}
\be  
S= \plm \int\!d^3x\, \sqrt{g}\,\Big( R - \, m^2 \, V [g_{\m\n}] \Big) \, .
\label{eq:act} 
\ee
A step forward in taming the zoo of possibilities was made in a series of
papers~\cite{uscan,usweak,uslong} through the nonperturbative construction of the most general
theories with five propagating degrees of freedom (DoF), characteristic of a massive
graviton.\footnote{See~\cite{Soloviev:2013mia} for a alternative analysis using Kuchar's Hamiltonian
  formalism.}  Besides its theoretical interest, the main phenomenological goal is to investigate
whether a modification of gravity at large distances and a massive graviton can be realized in a
consistent way and in agreement with the wealth of observational tests of gravity, from the smallest
(submillimiter) to largest (cosmological) scales.  Clearly, one of the crucial tests for a theory of
gravity is the existence of a realistic FRW cosmological evolution, which we address later.

The key tool for our analysis is the Hamiltonian formalism, which we will use to classify the
various potentials $V$ according to the number of degrees of freedom (DoF) that propagate.  In GR,
where $V=0$, among the ten components of the metric $g_{\mu\nu}$, diffeomorphism invariance gets rid
of eight of them.  As soon as we add extra nonderivative terms, diffeomorphism invariance is
broken. Diffeomorphism invariance can be restored by introducing a set of suitable Stuckelberg
fields, see for instance~\cite{HGS,dub,uslong}; of course such procedure does not change the number
of DoF. The action we consider can be obtained by choosing a suitable gauge (unitary gauge) where
the Stuckelberg fields are trivial; thus, all such a theories have a preferred frame.

On general grounds, once diffeomorphism invariance is broken by nonderivative interactions, one
expects six DoF, in contrast with the fact that at the linearized level a massive spin two particle
on Minkowski has 5 DoF~\cite{Fierz}. Indeed this was the conclusion reached in the past by Boulware
and Deser (BD)~\cite{BD}, by studying the nonlinear generalizations of a Lorentz invariant
Pauli-Fierz graviton mass term.  The mismatch between perturbative and nonperturbative number of
DoF is problematic because it is a signal of strong coupling. Moreover, the missing sixth mode is
a ghost on a Minkowski background.  Nevertheless, there exist particular choices of the potential
$V$ where this counting has to be refined and less than six DoF are nonperturbatively present. This
helps to get rid of the BD ghost and modifies also the phenomenology of these theories.
  
According to the Hamiltonian analysis \`a la Dirac, once we determine all first class (FC) and
second class (SC) constraints~\cite{Henn}, the number ($\#$) of DoF is given by
\be\label{dof}
\#\,\text{DoF}=10-\frac12 \# \, \text{SC}  - \# \, \text{FC}
\, .
\ee
The case of five DoF was discussed in detail in~\cite{uscan,uslong}, together with the
phenomenological~\cite{usweak} and cosmological~\cite{uscosm} consequences.

Among the various theories, one finds the very special case in which Lorentz symmetry is present
around Minkowski background~\cite{Gabadadze:2011}, that avoids the BD ghost~\cite{GF}, and is almost
unique~\cite{uscan}; see \cite{deRham} for a complete review on the subject.  This theory however is
phenomenologically not very successful: denoting with $m$ the graviton mass scale, the energy cutoff
$\Lambda_3=(m^2 M_{pl})^{1/3}$ is too low as already predicted in~\cite{HGS}; the theory is
classically strongly coupled in the solar system~\cite{Vainshtein,HGS} and even the computation of
the static potential in the vicinity of the earth is problematic due to quantum
corrections~\cite{HGS,pad,Brouzakis}.  Cosmology is also definitely troublesome: spatially flat homogenous
Friedmann-Robertson-Walker (FRW) solutions simply do not exist~\cite{DAmico} in the unitary gauge
and even allowing for open FRW solutions~\cite{open} strong coupling~\cite{tasinato} and ghostlike
instabilities~\cite{defelice} develop.  Another issue is the existence of acausal superluminal
modes~\cite{Deser}.  In the bigravity formulation~\cite{DAM1,PRLus, myproc, spher, energy, spher1}
FRW homogenous solutions do exist~\cite{cosm}, however cosmological perturbations turn out to be
strongly coupled~\cite{cosmpert}.

On the other hand, things get better if one gives up Lorentz invariance in the gravitational sector
and requires only rotational invariance~\cite{Rubakov,dub,usweak}. Within the general class of
theories which propagate five DoF found in~\cite{uscan,uslong}, in the Lorentz breaking (LB) case
most of the theories have a much safer cutoff $\Lambda_2 =(mM_{Pl})^{1/2}\gg
\Lambda_3$~\cite{usweak}, which is the maximal cutoff that one may obtain.  They also avoid all of
the phenomenological difficulties mentioned above~\cite{usweak,uscosm}.

In the present paper we complete the analysis started in~\cite{uscan,uslong} by considering all $V$
which respect rotational invariance, and classify them according to:
\begin{itemize}
\item the number of propagating DoF; 
\item the possibility of a viable FRW cosmology;  
\item the presence of strong coupling.
\end{itemize}

The outline of the paper is the following. In sect.~\ref{can}, by using Hamiltonian analysis, we
find for each $V$ the number of DoF ($\#$DoF).  In sect.~\ref{pert}, we compare the $\#$DoF found by
canonical analysis to the $\#$DoF computed using perturbation theory around Minkowski space; as a
result we can determine when strong coupling is present. In section 4 we study when a generic $V$
admits a FRW homogeneous solution in the unitary gauge that represents the reference background for
our expanding Universe.  In appendix~\ref{app:time} we extend our finding to potentials with an
explicit time dependence.  Our conclusions are given in section~\ref{conc}.

\section{Hamiltonian Analysis}
\label{can}

The standard Arnowitt-Deser-Misner splitting~\cite{ADM} of spacetime leads to the following
parametrization of the metric in terms of lapse $N$, shifts $N^i$ and spatial metric $\g_{ij}$:
\be 
g_{\mu \nu} = \begin{pmatrix} - N^2 + N^i N^j \g_{ij} &\;\; \g_{ij}N^j \\ 
\g_{ij}N^i  & \g_{ij} \end{pmatrix}. 
\ee 
The potential $V(g_{\m\n})$ is thus regarded as a function of  $N$, $N^i$ and $\g_{ij}$.

It is also useful to define 
\be
\V[N,N^i,\g_{ij}] \equiv 
m^2 \, N \, \sqrt{\g} \,V[N,N^i,\g_{ij}]\,, 
\ee 
with $\g=\det\g_{ij}$, and to write the Hamiltonian as
\be H=\int \!d^3x\,\Big[ {\cal
    H}_A(t,\vec x)\,N^A(t,\vec x)+\V(t,\vec x)\Big] \,, 
\ee
where we collected lapse and shifts in $N^A\equiv(N,\,N^i)$, with $A=0,1,2,3$, and the first piece
is the standard GR Hamiltonian.

Exactly as in GR, the lapse and the shifts appear in the Lagrangian with no time derivatives, so their
momenta vanish and lead to the four \emph{primary constraints}
\be
\Pi_A=\frac{\partial H}{\partial N^A}\approx0 \,, \qquad A=0,1,2,3.
\ee
These can be enforced by a set of four Lagrange multipliers $\lambda^A$ in the \emph{total
  Hamiltonian}
\be
H_T=H+\int \!d^3x\,\lambda^A(t,\vec x)\,{\Pi}_A(t,\vec x) \equiv H+
\lambda^A\cdot {\Pi}_A \, .
\ee
The time evolution of any function $F$ of $\g_{ij}$, $N^A$ or their momenta is given by the
Poisson bracket with $H_T$
\be
\frac{d F(t,\vec{x})}{dt} \equiv \Big\{ F(t,\vec{x}), H_T(t) \Big\} 
= \Big\{ F(t,\vec{x}), H(t) \Big\}
+ \int \!d^3 y \; \lambda^A(t,\vec{y})
\Big\{F(t,\vec{x}), \Pi_A(t,\vec{y}) \Big\} \, .
\ee
To avoid excessive cluttering, in the following we will mostly omit the time dependence of the
fields. If not stated explicitly, they are evaluated at the same time $t$.

The conservation in time of the primary constraints leads to four \emph{secondary}
constraints
\be
\S_A(\vec x)={\cal H}_A(\vec x)+\V_A(\vec x) \, \approx 0\,, \qquad A=0,1,2,3\,,
\ee
where $\V_A \equiv \de \V/\de N^A$.  Imposing again the conservation of the four secondary
constraints, leads to the \emph{tertiary} conditions
\be
 \T_A (\vec x)=\{\S_A(\vec x), \, H \}+\lambda^B(\vec x)
 \,\V_{AB}(\vec x) \approx 0 ,\,\qquad
 \V_{AB}\equiv\frac{\partial^2 \V}{\partial N^A\partial N^B} \, .
\label{ter}
\ee
The nature of these conditions, i.e.\ whether they are constraints or  determine some of the Lagrange
multipliers, depends on the rank of the Hessian of $\V$ with respect to $N^A=(N,N^i)$,
\be
 r = \text{Rank}\left|\V_{AB}\right| \, .
\ee 
The value of $r$ ranges between zero and four, the dimension of spacetime.


\subsection{$\mathbf{r=4}$: 6 DoF}
\label{sec:r=4}

If $r=4$, we can determine all four Lagrangian multipliers from~(\ref{ter}). All constraints are
consistent with the time evolution and the analysis stops here.  Thus, we end up with a total number
of DoF
\be
\# \text{DoF}=\frac{20-4 (\Pi_A)-4 (\S_A)}{2}=6 \, .
\ee
In other words, in the general case in which $\det|\V_{AB}| \neq 0$, we have $4\, (\Pi_A)+ 4\,
(\S_A)=8$ constraints, for a total of $6$ propagating DoF.  Technically, these 8 constraints are all
second class, being the ${\rm Rank}|\{ \Pi_A,\,\S_B\}|={\rm Rank}|\V_{AB}|=4$.  As a result, no
residual gauge invariance is present.  When the action is Lorentz invariant around a Minkowski
background, the six DoF must be organized in a massive spin two (5 DoF) representation plus a scalar
(1 DoF).  This is the Boulware-Deser result, valid for a generic potential.  The extra scalar, the
so called Boulware-Deser sixth mode~\cite{BD}, turns out to be a ghost around Minkowski space time,
rendering these generic theories hardly viable. 

It has to be stressed that the ghost can be absent around a FRW background, see section~6
in~\cite{diego}. As shown in that work, no ghost is present at any momentum if some conditions for
the graviton mass terms hold, $m_1^2 >0$ and $0<m_0^2< 6 H^2$ (see below section~\ref{pert} for the
notation). Moreover absence of tachyonic instabilities can also be fulfilled by further
conditions. The relative constraints on the potential may lead to an interesting scenario and we
leave it for a separate complete study.

In any case, a first result is that a necessary condition to have a theory with less than six
propagating DoF is that $r \equiv\text{Rank}|\V_{AB}|<4$ (see also~\cite{uscan}).

 
\subsection{$\mathbf{r<4}$: General analysis}
\label{sec:generalr<4}

Let us describe here in generality the hamiltonian analysis for $r<4$, and later specialize to the
various  cases $r=3,2,1,0$.  For $r < 4$, the matrix $\V_{AB}$ has $r$ non null eigenvectors,
denoted by $E^A_n$ with $n=1, \dots, r$, and $4-r$ null eigenvectors denoted by $\chi^A_\alpha$,
\be
\V_{AB}\,\chi^B_\alpha=0\,,\qquad \alpha=1,\dots,4-r\,. 
\ee
It is useful to decompose the Lagrange multipliers along those eigenvectors,
\be
\lambda^A=\sum_{\alpha=1}^{4-r}\,z_\alpha\,
\chi_{\alpha}^A+\sum_{n=1}^{r}\, d_n \, E^A_n \, ,
\label{proj}
\ee
effectively trading the 4 Lagrange multipliers $\l^A$ for the coefficients $z_\a$ and $d_n$.

\pagebreak[3]

Of the four original Lagrange multipliers, the $r$ components along $E^A_n$ are determined by
the tertiary condition (\ref{ter}):
\be
d_n= \frac{E^A_n\{\S_A,H\}}{E^A_n\,\V_{AB}\,E^B_n}   \, ,\qquad n=1,\dots, r \, . 
\ee
On the other hand, the projection of he conditions (\ref{ter}) along the null directions
$\chi_\alpha^A$ unveils $4-r$ genuine \emph{tertiary constraints}
\be
\T_\alpha \equiv \chi_\alpha^A\,\{\S_A,\,H\}\approx 0\,,\qquad
\alpha=1,\dots,4-r\, .
\ee
Indeed, no Lagrange multiplier is involved here.

We have also to impose the conservation in time of these new constraints, which leads to the
conditions
\be
\begin{split}
\Q_\alpha(\vec x) = \{ \T_\alpha(\vec x), \, H\} 
+\int d^3y\, \Bigg [\, & \sum_{n=1}^r \, d_n(\vec y) \, 
\{ \T_\alpha(\vec x) , \, \Pi_A (\vec y) \} \,     E^A_n(\vec y)  
\\& - \sum_{\beta=1}^{4-r} \, \theta_{\alpha\beta}(\vec x,\vec y)  \,
z_\beta(\vec y) \Bigg] \;\approx \;0 \, ,
\end{split}
\label{quat}
\ee
where the matrix
$\theta_{\alpha \beta}$ is defined as
\be\label{theta}
{\cal \theta}_{\alpha \beta}(\vec x,\,\vec y)
\equiv
\chi_{\alpha}^A(\vec x) \,\{\S_A(\vec x),\,\S_B(\vec
y)\}\,\chi_{\beta}^B(\vec y) \, .
\ee
The condition (\ref{quat}) consists in $4-r$ linear equations for the remaining $4-r$ Lagrange
multipliers $z_\alpha$.  Hence, the number of DoF crucially depends on how many of them can be
determined, i.e.\ on the rank of $\theta_{\a\b}$
\be
s\equiv\text{Rank}\, |{\cal \theta}_{\alpha \beta}| \,.
\ee
If $s=4-r$, then all the remaining Lagrange multipliers are determined and the procedure which
enforces the consistency of constraints with time evolution ends.  On the other hand if $s<4-r$ some
of the $z_\alpha$ are not determined and one has \emph{$4-r-s$ new quaternary constraints} $\Q_\a$,
which further reduce the number of DoF.

Altogether so far one has $16-2\,r-s$ constraints, counting $4\, (\Pi_A)+ 4 \, (\S_A) +( 4-r)\,
({\cal T_\alpha})+(4-r-\,s) \, (\Q_\a)$, and the number of DoF is at this point
\be
 \#\,\D\le \frac{20- (16-2r-s)}{2}= 2+r+\frac{s}{2} \, ,  \qquad 0
 \leq r \leq 4 \, , \qquad 0 \leq s \leq 4-r
 \, .
\label{DoF}
\ee
Maximizing $s$, for  fixed $r$, we have the  following upper bound
\be
\#\,\D\le   4+\frac{r}{2}  \, ,  \qquad 0\leq r \leq 4 \, .
\ee

Once more, in order to know how far one can go, one has to check the conservation of the quaternary
constraints, that reads
\be
\F_\alpha (\vec x)= \{ \Q_\alpha(\vec x), \, H_T \} =  \{ \Q_\alpha(\vec x), \, H\} +\int \!d^3y\,
\frac{\de \Q_\a(\vec x)}{\de N^A(\vec y)}  \left( \sum_\b
\chi^A_\b \, z_\b + \sum_n  E^A_n \, d_n  \right)(\vec y) \approx 0 \, .
\label{fifth1}
\ee
Setting 
\be
A_{\a \b}(\vec x,\,\vec y) =  \frac{\de \Q_\a(\vec x)}{\de N^A(\vec
  y)} \chi^A_\b(\vec y) \,,
\ee
if the matrix $A_{\a\b}$ is invertible, then (\ref{fifth1})\ does not give rise to new
constraints but simply determines the remaining Lagrange multipliers as
\be
z_\a \propto - \sum_\b A^{-1}_{\a \b} \left( \{ \Q_\b, \, H\} + \sum_n \,d_n\,E^A_n
\, \frac{\de \Q_\a}{\de N^A}  \right) \, .
\label{lastll} 
\ee
In this case, the procedure ends here and the number of DoF saturates the bound in (\ref{DoF}).
However, again this is only the maximal number. In fact, if some of the $z_\a$ are not determined, more
steps are necessary and the net effect is to reduce the number of DoF further.  In general, also
first class constraints may be present corresponding to residual gauge invariances, but again, this
implies a further reduction of the number of DoF.  In the above discussion we have also ignored the
exceptional cases where some constraints are accidentally trivial, e.g. $0 \approx0$.

It is important to remark that due to the nontrivial dependence on $\vec x, \,\vec y$, the matrix
$\theta_{\alpha \beta}(\vec x, \,\vec y)$ is not necessarily antisymmetric and its rank $s$ is not
always even. Thus, for $s$ {\it odd} one concludes that an \emph{half integer} number of DoF is
present.  This is a peculiar phenomenon which arises in classical field theories (infinite
dimensional Hamiltonian systems). It is briefly discussed in general terms in~\cite{1/2dof} and for
Horava-Lifshits gravity in~\cite{henneaux}.  To our knowledge, no general analysis on the nature of
such half DoF is present in the literature.  We leave the matter for a future investigation.  From
our general analysis, the only viable case with an half integer number of DoF is the one with
5+$\frac{1}{2}$ DoF, see section~\ref{sec:r=3}. 

We recap the steps that are required to compute the number of propagating DoF for a given deforming
potential $\V$:
\begin{enumerate}
\item Compute the rank $r$ of the hessian matrix $\| \V_{AB}\|$ ($4 \times 4$ matrix).  
\item Compute the null eigenvectors $\chi^A_\a$ of the matrix $\V_{AB}$.
\item Determine secondary constraints $\S_A=\Hc_A+\V_A$.
\item Compute the rank $s$ of the matrix
$\| \chi^A_\a\,\{\S_A,\,\S_B\}\,\chi^B_\b \|$ ($4-r \times 4-r $ matrix).
\item Plug the numbers in the formula
$\# \,\D\leq 2+r+s/2$.
\end{enumerate}

In the following sections we discuss separately the cases relative to different values of $r$ and
$s$. The results of this analysis are summarized in Table~\ref{tab:dof}, where the maximal number of
DoF resulting from the canonical analysis is shown for different values of $r$ and $s$. We also
report whether the resulting theory can be built by respecting rotations (fourth column), and
whether it can be realized at all with some explicit form of the potential (last column), as we find
by direct inspection in the forthcoming sections.

\begin{table}[t]
  \centering
  \begin{tabular}{|c|c|c|c|c|}\hline\hline
    $r$ =Rank$|\V_{AB}|$   & $s$ =Rank$|\theta_{\a\b}|$   & $\#$DoF $\leq$& Rotations? & Realized?\\ \hline\hline
    \bf 4 &\bf  0   &\bf  6 & $\mathbf\surd$ &\bf  Yes \\ \hline   \hline
    \bf 3 & \bf 0   &\bf  5 &\bf $\surd$ &\bf Yes \\ \hline   
   \bf  3 & \bf 1   &\bf  5+$\frac{1}{2}$ & \bf $\surd$ &\bf Yes \\ \hline   \hline
    2 & 0   & 4 & $\times$& No\\\hline   
    2 & 1   & 4+$\frac{1}{2}$ & $\times$& No\\\hline   
    2 & 2   & 5 & $\times$& No\\ \hline   \hline
   \bf  1 &\bf  0  & \bf 3 &\bf  $\surd$ &\bf Yes \\ \hline   
    1 & 1   & 3+$\frac{1}{2}$ & $\times$& No\\ \hline   
    1 & 2   & 4 & $\times$& No\\ \hline   
    1 & 3   & 4+ $\frac{1}{2}$ & $\surd$& No\\ \hline  \hline
  \bf   0 &\bf 0   &\bf 2 &\bf $\surd$&\bf Yes \\ \hline   
    0 & 1   & 2+ $\frac{1}{2}$ & $\surd$& No\\ \hline   
   \bf  0 &\bf  2  &\bf  3 &\bf  $\surd$&\bf  Yes \\ \hline   
    0 & 3   & 3+$\frac{1}{2}$ & $\surd$& No\\ \hline   
    0 &4  &4& $\surd$& No\\ \hline   \hline
  \end{tabular}
  \caption{%
    Deforming potentials classified according to the rank $r$ of the Hessian and the rank $s$ of the
    matrix $\theta$. The number of DoF is obtained from eq.~(\ref{DoF}). The cases consistent with the 
    canonical algebra are highlighted as bold and marked as {\it Realized}.  Notice that a non integer number 
    of DoF can possibly appear with unbroken rotations only in the case $r=3$ and $s=1$, 
    namely 5+$\frac{1}{2}$ DoF.%
  }
  \label{tab:dof}
\end{table}

A first outcome of the analysis is that massive deformations of gravity with 5 DoF exist only in two
cases: $r=3$, $s=0$ or $r=2$, $s=2$. The first was discussed in full depth in
\cite{uslong,usweak,uscan} where all the rotational invariant potentials of this class where
constructed.  Concerning the second case, we note that it is not possible to build potentials with
$r=2$ without breaking spatial rotations. Indeed, rotational invariance requires either one or at
least three non null eigenvectors of $\V_{AB}$.

In the following we consider potentials that are at least rotationally invariant on Minkowski space.
As a result, we are left only with the cases $r=4,3,1,0$, with a number of DoF between 6 and 2.

Remarkably, we can already exclude the models with 4 DoF, often invoked in the context of massive
gravity (see for instance~\cite{Deser-par}); they in principle could exist for $r=2$, $s=0$, but
only with broken rotational invariance.  The candidates with 4 DoF having $r=0$, $s=4$ are actually
not realized, as we will see in Section~\ref{sec:r=0}. Thus, we conclude that the only candidate
theories with 4\,DoF are to be searched as subcases of the 5\,DoF theories with $r=3$. This is
discussed in section~\ref{sec:4dof}; the high number and complexity of the required constraints
makes one doubt that such theories can actually be found.


\subsection{The case $\mathbf{r=3}$: massive gravity with 5 DoF}
\label{sec:r=3}

This case was fully analysed in~\cite{uscan, usweak, uslong} (see also \cite{Soloviev:2013mia} for a
similar approach). Here for completeness we recollect the main results.  The general potential $\V$
of massive gravity theories with five propagating DoF can be parametrized in terms of two arbitrary
functions of specific arguments, $\U[\K^{ij}\equiv\g^{ij}-\xi^i\,\xi^j]$ and $\E[\xi^i,\g^{ij}]$,
\be
  \V\equiv m^2\,\sqrt{\g}\,\Big(N\,\U +\E+\U_i\,\Q^i \Big)\,,
\ee
where  $\xi^i$ is defined implicitly by the first of the following equations
\be
  N^i=N\,\xi^i+\Q^i\,,\qquad \Q^i[\xi^i,\g^{ij}]\equiv-\,\U_{ij}^{-1}\,\E_j\,.
\nonumber
\ee
and where $\U_i=\de_{\xi^i}\U$ and $\U_{ij}=\de^2\U/\de \xi^i\de \xi^j$. The use of the variables
$\xi^i$ in place of the shifts $N^i$ makes also very transparent the canonical analysis, as recalled
in appendix~\ref{app:4dof} (see also~\cite{uslong}).  The function $\E$ is the bulk on-shell energy
(Hamiltonian) density of the system and it has to be non-negative.  We remark that, as shown in
appendix \ref{app:4dof}, a necessary condition to have 5 DoF, is to have $\E\neq 0$. Potentials with
$\E=0$ have six DoF.

Besides its purely theoretical interest, this result is also relevant from a phenomenological point
of view.  A large class of massive gravity theories that are ghost free on Minkowski space are
uncovered, whereas previously, the only known ghost free theory was the four parameter Lorentz
invariant (LI) theory found~\footnote{Also Zumino came up with a
similar model, see  
Brandeis Univ. 1970, Lectures On Elementary Particles And Quantum Field Theory, Vol. 2*, Cambridge,
Mass. 1970, 437-500.} in~\cite{Gabadadze:2011,GF}, which is a special case of our general
construction.\footnote{For instance, the minimal version of the dRGT LI massive gravity is obtained
  by taking $\U=(Tr[\K^{1/2}]-3)$ and $\E=(1-\xi^2)^{-1/2}$, that gives rise to their potential
  $(Tr[\sqrt{X}]-3)$ with $X^\mu_\nu=g^{\mu\rho}\,\eta_{\rho\nu}$. For the other operators in that
  theory the correspondence is not known so explicitly.}  When Lorentz symmetry is enforced, the
price to be paid is the impossibility of using perturbation theory in many physical important
situations like inside our solar system. Moreover, as effective theory, the cutoff is rather
low~\cite{HGS}, $\Lambda_3=(m^2M_{Pl} )^{1/3}\sim 10^3\,$Km when $m$ is taken to be of order of
today's Hubble scale. As a result, even the static potential between two masses at a distance
smaller than $10^3\,$Km is difficult to compute perturbatively~\cite{HGS,pad,Brouzakis}, in contrast
with short distance tests of Newton's force at submillimeter scale, see for instance~\cite{Geraci}.
In Lorentz breaking theories we are much better off from a phenomenologically point of view.  It
ought to be remarked that Lorentz symmetry we are discussing here only concerns the gravitational
sector and is not the same symmetry that enters in the formulation of the Einstein's equivalence
principle. As such, it is not subject to strong phenomenological constraints coming from high energy
physics.  Thus, we conclude that the viability of the theory is directly connected with the need to
have give up Lorentz symmetry in the gravitational sector, which is testable in the forthcoming
gravitational wave experiments.

The concrete phenomenology of the new class of Lorentz breaking theories is also rather promising,
as argued in~\cite{usweak}. From a perturbative point of view, exploiting the general expression of
$\V$, there exist remarkable relations among the various Lorentz breaking graviton masses.  At the
nonperturbative level, besides the absence of ghosts in the spectrum, it is of crucial importance to
be able to trust the theory up to the cutoff $\Lambda_2=(m M_{PL})^{1/2}\simeq (10^{-3}
\,\text{mm})^{-1}$, as the absence of strong nonlinearities (Vainshtein effect) around macroscopic
sources.  Such class of Lorentz breaking massive gravity theories is also a natural candidate for
dark energy provided its equation of state deviates from -1~\cite{uscosm}.


\subsection{The subcase of $\mathbf{r=3}$ for massive gravity with 4 DoF}
\label{sec:4dof}

In appendix \ref{app:4dof} we give the further conditions under which a potential with $r=3$
propagates only four DoF.  In comparison with the case of 5 DoF, two extra (differential) conditions
on the potential have to be imposed.  In the Dirac language, they correspond to the requirement that
the quinary and the senary constraints are independent from the lapse
\be
\partial_N\Q=\partial_N {\F}=0 \, .
\ee
These conditions restrict further the dependence on the ADM variables of the functions $\U$ and
$\E$.  However, due to their complexity, at present no solution is known, if any exists.  In this
sense no $V$ with 4 DoF is known.  As it will be discussed in section~\ref{pert}, around Minkowski
background only two or five DoF can propagate at linearized level; thus, even if a potential with
four DoF exists it will lead to strong coupling around flat space.

At linearized level, Lorentz-violating potentials which propagate 4 DoF (two tensor and two vector
modes) were analyzed even around a generic FRW background (see ref.~\cite{diego}). For instance, on
de Sitter background if the graviton mass is precisely $m^2 = 2H^2$, a fifth scalar mode disappears
from the linearized theory, leading to the so called partially massless (PM) theory~\cite{DW}.  The
absence of the helicity-0 mode at linearized level is related to the existence of a new scalar gauge
symmetry (a special combination of a linearized diff. and a conformal transformation).
Unfortunately, the helicity-0 mode reappears non-linearly; so, rather than being free from the
scalar mode, the theory is strongly coupled~\cite{DW1}.


\subsection{The case $\mathbf{r=1}$}
\label{sec:r=1}
When $r=1$ and rotations are preserved, the only possible form for $\V$ is  a function of
$\gamma_{ij}$ and $N$ with nonzero $N$ second derivative: 
\be
\V \equiv \V[N,\g]\,, \qquad\text{with}\qquad\V_{NN}\neq0\,.
\ee
Following the steps of section~\ref{sec:generalr<4}, we have
\begin{itemize}

\item The secondary constraints in this case are rather simple
\be
\S_0=\Hc + \V_N \approx 0 \, , \qquad \S_i=\Hc_i 
\approx 0 \, , \qquad i=1,2,3.
\ee
\item There are three null eigenvectors that can be chosen to be $\chi^A_i = \delta^A_i$.

\item The matrix $\theta_{\a \b}$ of (\ref{theta}) now vanishes when the constraints are used,
  namely
\be
\theta_{\alpha \beta}(\vec x,\vec y) \equiv \theta_{ij}(\vec x,\vec y) = \{ \Hc_i(\vec x) , \,
\Hc_j(\vec y) \}\propto \Hc_k
\approx 0  \, ,
\ee
where GR algebra has been used. Thus $s=0$.

\item $\#\, \D= 3$.

\end{itemize} 
The on-shell  bulk Hamiltonian is given by
\be
H_{|\text{on shell}}= \int\!d^3x\,\left(\V-\V_N\,N \right)(t,\vec x )
\, .
\ee


\subsection{The case $\mathbf{r=0}$}
\label{sec:r=0}

When $r=0$ and $\V$ is rotational invariant, the only possibility is that $\V$ is at most a linear
function in the lapse, hence we can write
\be
\V \equiv m^2\,\sqrt{\g}\,\Big(N \,{\bold  U}[\gamma_{ij}] + {\bold E}[\gamma_{ij}] \Big) .
\ee
The Hamiltonian analysis for specific examples in this class was already given in \cite{grisa}.

\vspace{0.3cm}
Consider first the case of generic ${\bold U}\neq 0$:
\begin{itemize}
\item The null eigenvectors of the hessian can be chosen to be $\chi^A_\a=\delta^A_\a$ with $\a=0,1,2,3$.
\item The secondary constraints are 
\be
\S_0=\Hc + m^2\,\sqrt{\g}\,{\bold U}  \approx 0  \, , \qquad \S_i=\Hc_i 
\approx 0 \,, \qquad i=1,2,3\,.
\ee
\item We calculate $\theta_{\a \b}=\{\S_{\a},\,\S_\b\}$ by using the same algebra of constraints of
  GR:
\bea
\theta_{00}  &=&[\Hc_i(\vec x)+\Hc_i(\vec y)]\partial_{x^i}\,\delta(\vec x-\vec y)\approx0, \nonumber\\
\theta_{0i}   &=& 
  \sqrt{\g}\,\partial_{x^i}\left({\bold U}_\g+\frac{\bold U}{2}\right)\delta(\vec x-\vec y)+ 
\sqrt{\g}\left(\g_{ia}\,\partial_{\g_{ab} }\,{\bold U}\right)\,\partial_{x^b}\delta(\vec x-\vec y)\neq 0,\nonumber\\ 
\theta_{ij} &=& \left(\Hc_j(\vec x)\,\partial_{x^i}+\Hc_i(\vec y)\,\partial_{x^j}\right)\,\delta(\vec x-\vec y)\approx 0 \, ,
\label{eq:theta0}
\eea
so that clearly $s=2$.
\item As a result, $\#\, \D=3$.
\end{itemize} 

A subcase is also present, which is relevant to our analysis. We note that the equations
(\ref{quat}) for the Lagrangian multipliers $z_\alpha$ associated with the $\theta_{0i}$ given above
are linear differential equations in the spatial coordinates, of the form ${\it A}\partial_x
z+Bz=C$. A similar structure is found in Horava-Lifshitz gravity~\cite{henneaux}.  Being
\bea
{\it A}\propto \g_{ia}\,\partial_{\g_{ab} }{\bold U}\,,
\qquad
 {\it B}\propto \partial_{x^i}\left({\bold U}_\g+\frac{\bold U}{2}\right),
\eea
it is easy to show that there is a unique potential ${\bold U}$ (function of the rotational
invariants ${\rm Tr}[\g]$, ${\rm Tr}[\g^2]$, ${\rm Tr}[\g^3]$) such that ${\it A}$ and ${\it B}$
vanish automatically. This corresponds to a  cosmological constant, i.e.\ $ m^2\,{\bold
  U}=\Lambda$, or 
\be
\V \equiv \sqrt{\g}\,\Big(N \,\Lambda + m^2\,{\bold E}[\gamma_{ij}] \Big) \,.
\ee 

\begin{itemize}
\item In this case the Lagrange multipliers are not determined because even $\theta_{0i}$ vanish:
\be
\theta_{00} \propto\Hc_i\,\partial_i\,\delta \approx 0, \qquad
\theta_{0i}\propto \partial_i\Hc \approx 0,\qquad 
\theta_{ij} \propto \Hc_i\,\partial_j\,\delta\approx 0 \, ,
\ee
thus $s=0$.
\item As a result, $\#\,\D=2$.
\end{itemize} 
Note, since still $\V$ contains $m^2{\bold E}[\g_{ij}]\neq0$, the tertiary constraints $\T_\alpha$
do not vanish, and the Lagrange multipliers are determined at the level of quinary conditions
$\F$. This corresponds to a case of 2 \D\ but broken diffeomorphisms.  This situation contains
trivially also the case ${\bold U}= 0$. Clearly instead for ${\bold E}=0$ and $\V=\Lambda
N\sqrt{\g}$ the result is GR with cosmological constant and 2\,DoF are present, with unbroken gauge
invariance.

In all these cases the on-shell bulk Hamiltonian is given by
\be
H_{|\text{on shell}}=  m^2\int \! d^3x\,\sqrt{\g}\,{\bold E}[\g_{ij}](t,\vec x ) \, .
\ee
%

 
\section{Perturbations around Minkowski}
\label{pert} 

Consider now the perturbative expansion around flat space. Setting $g_{\mu \nu} = \eta_{\mu \nu} +
h_{\mu \nu}$, expanding the action (\ref{eq:act}) at the quadratic order in $h$ one gets
\be
S= \int \!d^4 x\, 
\left[ L_{(2)} + \frac{1}{2} \Big(
    m_0^2 \, h_{00}^2 + 2\, m_1^2 \, h_{0i}^2 -m_2^2 \, h_{ij}^2
    + m_3^2 \, h_{ii}^2 -2\,  m_4^2 \, h_{00} h_{ii} \Big) \right]  ,
\label{masses}
\ee
where $ L_{(2)}$ is the standard quadratic Lagrangian for a massless spin 2 particle in Minkowski
space; for $V$ we have only imposed rotational invariance.
The physical consequences of the quadratic action (\ref{masses}) were first discussed
in~\cite{Rubakov}.  In our case, the various masses can be computed explicitly from the potential.
Preliminarily, one has to impose that Minkowski space is a consistent background, i.e.\ that $g_{\mu
  \nu} = \eta_{\mu \nu}$ is a solution of the equations of motion; this is equivalent to
\be
\bar \V_N =0 \, , \qquad 
\bar \V_\g  =0 \, .
\label{flat}
\ee 
The bar indicates that expressions are evaluated on Minkowski space, where we define $\V_\g$ by
$\de \V/\de \gamma^{ij} \equiv \V_\g \, \gamma_{ij}\,.$
Using the conditions (\ref{flat}), we find that
\be
\begin{split}
m_0^2 =\left.-\frac{1}{4} \frac{\de^2  \V}{\de N^2}\right |_{\eta} \, ,\qquad
m_1^2 =\left.  - \frac{1}{2}  \frac{\de^2  \V}{\de N^i \de
  N^i}\right |_{\eta}    \, .
\end{split}
\label{massval}
\ee
The expressions for $m^2_{2,\,3,\,4}$ are not particularly illuminating and will be omitted. In
general, the following conclusions can be drawn~\cite{Rubakov,dub,PRLus,diego}:
\begin{itemize}
\item For $m^2_{0,\,1}\neq0$ we have 6 perturbative DoF with one scalar as a ghost around Minkowski
  background. At most, 6 healthy modes can be obtained around FRW spaces if $m_1^2 >0$ and $0<m_0^2<
  6 H^2$, plus other conditions to avoid tachyonic instabilities~\cite{diego}.
\item For $m_0^2=0$, $m^2_{1}\neq0$ we have 5 perturbative DoF.
\item For $m_0^2\neq 0$, $m^2_{1}= 0$ we have 2 perturbative DoF.
\item For $m^2_{0,\,1}= 0$ we have 2 perturbative DoF.
\end{itemize}

In the summary Table~\ref{tab:summa} we compare the number of perturbative DoF around Minkowski
space found in the present section with the corresponding number found in the previous section by
using the nonperturbative and background independent analysis.  If the two numbers differ, the
propagation of the missing DoF(s) have to show up at higher orders in the perturbative expansion, or
around non-Minkowski backgrounds.  In either case, this is a manifestation of strong coupling around
Minkowski spacetime.

 
\section{Cosmology}
\label{cosmo}
In this section we analyze the conditions under which the potential $\V$ admits a FRW background
solution.  We take for FRW metric the following diagonal form
\be
g_{00}=-N^2\, , \quad g_{0i}=0\rightarrow N^i=0\,,\quad
g_{ij}=\g_{ij}=a^2\,\delta_{ij} \, .
\label{FRW}
\ee
Notice that this is the most general ansatz with maximally symmetric $t=$const
hypersurfaces. Equivalently, the reference frame where the universe is homogenous is the very same
frame of the unitary gauge~\cite{uscosm}.  For simplicity we have also set the spatial curvature to
zero.

Due to the diagonal form of the FRW metric, its existence  probes the
functional dependence of $\V$ with respect to $N$ and $\g_{ij}$ only, no constraints on the $N^i$
dependence can be obtained.  The effect of $\V$ is equivalent to the presence in the Einstein
equations of an effective energy momentum tensor (EMT) $\T_{\mu \nu}$ defined by 
%
\be
  \delta \, \int \!d^4x \;  \V \equiv  \frac{1}{2} \int \! d^4x
  \;\sqrt{g}\,\T_{\mu \nu}  \, \delta g^{\mu \nu} 
\ee
and given by 
\be
\T_{\mu \nu} =\frac{2 }{\sqrt{g}} \, \frac{\de \V}{\de g^{\mu \nu}} \,.
 \ee
Specializing to the FRW background, the effective EMT reads
\be
\T_{00} = \frac{N^2}{\gamma^{1/2}}  \V_N \, , \qquad
\T_{0i}=0 \, , \qquad \T_{ij} =
\frac{2}{N \gamma^{1/2}} \,  \gamma_{ij}\, \V_\g \, ,
\ee
where we denote $\V_N\equiv\de \V /\de N $ and we have used the fact that on FRW background ${\de
  \V}/{\de \gamma^{ij}}$ is proportional to $\gamma_{ij}$ by defining $\V_\g$ through $\de \V/\de
\gamma^{ij}\equiv \V_\g \, \gamma_{ij}$. (For instance, for $\V=\g^n $, with $\g={\rm Det}[\g_{ij}]$,
we have $ \V_\g =-n\g^n$).  We retain here the explicit $N$ dependence of $g_{\mu \nu}$; indeed,
besides being instrumental in exploiting constraints on the functional dependence of $\V$ on $N$ and
$\g_{ij}\propto a$, in general it cannot be gauged away.

The gravitational fluid has energy and pressure densities and effective equation of state given by
\be
\label{eq:fluid}
\rho_{\text{eff}}=\frac{\V_N }{\, \g^{1/2}}\,,\quad  \qquad  p_{\text{eff}}=\frac{2 \, \V_\g}{N\,
  \g^{1/2}}\,,\qquad  w_{\text{eff}}=\frac{2 \, \V_\g}{N \, \V_N} \, .
\ee

Because of the Bianchi identities, $\T_{\mu \nu}$ must be covariantly conserved. This requires
$\partial_t \rho_{\text{eff}}=-3\,\frac{\dot a}{a}\,(\rho_{\text{eff}}+p_{\text{eff}})$,
which  leads to the  condition
\be\label{bianchi}
\dot N\,\V_{NN}
-6\,{\frac{\dot a}{a}}
\left(\V_{N\g}-\frac{\V_{\g}}{N}\right)=0 \, .
\ee
%
Notice that
$\partial_t \V=\g_{ij}\, \dot\g^{ij}\,\V_\g +\dot N\,\V_N=-\,\frac{6\,\dot a}{a}\,\V_\g +\dot
N\,\V_N$.  In general eq.\ (\ref{bianchi})\ is a differential equation which dictates the dynamics
of $N$, as can be seen by solving for $\dot{N}$. In this case $N$ is dynamically determined (and
cannot be eliminated by a choice of time). Then, the Friedmann equation determines the time
dependence of the Hubble parameter, and results in a well-behaved cosmology in the presence of the
effective fluid~\ref{eq:fluid}.

However, looking at the classification of admissible potentials from section~\ref{can}, we see that
in most cases the situation is crucially different.  Except in the case $r=1$, in all cases ($r=3$
or $0$) on FRW background where one has $N^i=\xi^i=0$, the potential is a linear function of the
lapse $N$. Thus $\V_{NN}=0$ and eq.\ (\ref{bianchi})\ has to hold in the form (we consider $\dot
a\neq 0$ for a realistic cosmology)
\be
\V_{NN}=0\, ,\qquad  \left(\V_{N\g}-\frac{\V_{\g}}{N}\right)=0 \, .
\label{bianchis}
\ee
Parametrizing $\V$, we find that the Bianchi condition constrains only the $N$-independent part:
\be
\label{eq:condcosm}
\V\equiv m^2\,\sqrt{\g}\,\Big(N{\cal A} +{\cal B}\Big) 
\qquad \Rightarrow\qquad
 \Big(\sqrt{\g}\,{\cal B}\Big)_{\g}=\sqrt{\g} \left.\left({\cal B}_\g-\frac{\cal B}{2}\right)\right|_{FRW} =0\,,
\ee
where, as in the cases of section~\ref{can}, ${\cal A} $ and ${\cal B}$ are generic functions (of
the spatial metric, on FRW).  Note that $\cal A$ as well as $N$ drop out of the Bianchi condition,
so that $N$ is now left undetermined by the background equations.  More importantly, we see that in
general the Bianchi condition ends up in an algebraic constraint on the scale factor $a$, which is
incompatible with a realistic cosmology. Thus, the only possibility is that some specific form of
$\cal B$ is chosen so that ${\cal B}_\g-\frac{\cal B}{2}=0$ identically on FRW. As an example,
${\cal B}= 3\,\tr[\g^2]-\tr[\g]$ has this property. This is the condition to have a realistic FRW
cosmology, and will apply to the cases of sections~\ref{sec:r=3}, \ref{sec:4dof} and \ref{sec:r=0}.

Density pressure and equation of state of the gravitational fluid now take the form
\be
\rho_{\text{eff}}=m^2\,{\cal A} \,,\qquad p_{\text{eff}}=2\,m^2\left(
  {\cal A}_\g -\frac{1}{2}\,{\cal A}\right)
\qquad \Rightarrow \qquad 
w_{\text{eff}}=-1+\frac{2\,{\cal A}_\g}{{\cal A}} \, .
\ee
We note that they do not depend on the function ${\cal B}$.\footnote{As a check, for a
  cosmological constant,$\V\propto N\g^{1/2}$, we have ${\cal A}=1$ and ${\cal A}_\g=0$ giving
  exactly $w_{\text{eff}}=-1$.} This can be explained by observing that only the function ${\cal B}$
breaks time reparametrizations, in the potential.  The function ${\cal A}$ appears in the
combination $N{\cal A}$ and has the same structure of the Hamiltonian constraint in GR.  Thus it
cancels out from the Bianchi condition (\ref{bianchi}) which is exactly the constraint related to
the (breaking of) time reparametrizations.  In other words, the part of ${\cal T}_{\mu\nu}$
containing ${\cal A}$ is automatically conserved, while the remaining part containing ${\cal B}$ has
to be conserved by itself.  We stress that from the existence of a FRW background nothing can be
said on the $N^i$ dependence of $\V$. Thus, a FRW solution would exist when the potential has the
general structure
\be
\V=m^2\,\sqrt{\g}\,\bigg(N{\cal A}\Big[\g, \,N^i f_{1}[\g,\,N,\,N^k] \Big]+{\cal B}\Big[\g, \,N^i f_2[\g,\,N,\,N^k] \Big] +N^i\g_{ij}N^{j}\,f_3[\g,\,N,\,N^k]\bigg)\, ,
\ee
with $f_i$ generic functions and where ${\cal B}$ must again be chosen such that ${\cal B}_\g=0$ on
FRW.

\medskip

We can now put together the results above with the analysis of the previous sections, and spell out
the potentials which exist and admit a consistent FRW background:
\begin{itemize}
\item[({\bf a})] 6 DoF with $r=4$. Two tensors, two vectors and two scalar modes are present, of
  which one is a ghost around Minkowski spacetime.  As recalled in section~\ref{pert}, the ghost can
  be absent around FRW backgrounds if $H'<0$~\cite{diego,uscosm}.
\item[({\bf b})]
 5 DoF with $r=3,\,s=0$. Here we have
\bea
 \V&=& m^2\,\sqrt{\g}\,\Big(N\,\U\!\left[\g^{ij}-\xi^i\xi^j\right]+ \E\left[\g^{ij},\xi^i\right]+\U_{\xi^i}\Q^i\Big) \;\; \text{and}\;\; w_{\text{eff}}=-1+\frac{2\,\U_\g}{\U},\ \ 
\eea
where we recall that $ N^i=N\,\xi^i+\Q^i$ and $\Q^i=-\|\U_{\xi^i\xi^j} \|^{-1}\E_{\xi^j}$.

The potential is such that (\ref{bianchi}) is solved in the form (\ref{bianchis}). The existence of
a nontrivial FRW solution requires $\E_\g=\E/2$. This, if combined with the condition for the
existence of a strict Minkowski background (\ref{flat}), predicts that the $m_1^2$ in (\ref{masses})
is zero. This leads to strong coupling in the vector and scalar sector of perturbations around the
strict Minkowski background. The same conclusion is reached for a strict de Sitter space, and the
only healthy possibility is to deviate from a de Sitter phase~\cite{uscosm}.

\item[({\bf c})] 3 DoF with $r=1,\,s=0$. In this case, generically (\ref{bianchi}) can be solved for
  $\dot{N}$ and a FRW solution exists.  The dependence of $N$ on $a$ strongly depends on the
  explicit form of $\V$. On Minkowski spacetime strong coupling of gravitational perturbations is
  present, see Table~\ref{tab:summa}.  Instead, by using the results of~\cite{diego}, we see that
  around FRW 3 DoF are present in the linearized theory, thus in agreement with the 3
  nonperturbative DoF as predicted by the canonical analysis.

\item[({\bf d})] 3 DoF with $r=0,\,s=2$. Here 
\bea
 \V &=& m^2\,\sqrt{\g}\,\Big(N\,{\bold U}[\g]+\,{\bold E}[\g] \Big)
 \qquad \text{and} \quad w_{\text{eff}}=-1+\frac{2{\bold U}_\g}{\bold U}\,.
\eea
The Bianchi condition is realised in the form (\ref{bianchis}), thus $\left. {\bold E}_\g -{\bold
    E}/2\right |_{FRW}=0$ must be satisfied.  For the perturbations, the same considerations hold as
in case ({\bf c}) .

\item[({\bf e})] 2 DoF with $r=0,\,s=0$. Here
\be
\V=\sqrt{\g}\,\Big(N\,\Lambda  +m^2\,{\bold E}[\g]\Big)
 \qquad \text{and} \quad w_{\text{eff}}=-1\,.
\ee
Even in this case the conservation of the effective EMT is realised in the form (\ref{bianchis})\
and thus one needs $ {\bold E}_\g -{\bold E}/2|_{FRW}=0$. However
$\rho_{\text{eff}}=-p_{\text{eff}}=\Lambda$, hence the effect on cosmology of this class of
potentials is indistinguishable from a plain cosmological constant.  Differences with GR may
appear in spherically symmetric Schwarzschild-like solutions.  While the matter is beyond the
scope of the present analysis, following~\cite{diego} we can anticipate that no vDVZ discontinuity
is found at the linearized order, both on Minkowski and on FRW backgrounds. Thus, GR is recovered
smoothly in the limit of small graviton mass.  The absence of discontinuity implies also the absence
of Vainshtein spatial strong coupling. As a result, these models could represent a new interesting
class of massive gravity theories.
\end{itemize} 
%
The results are summarized in table~\ref{tab:summa}.

\begin{table}[b]
  \footnotesize
  \small
  \renewcommand\arraystretch{1.3}
  \centering
  \begin{tabular}{|l||c|c|c|l|}
    \hline
    {\bf } & {\bf \!Nonpert.\!}& {\bf LB } &{\bf Pert.}& {\bf FRW }   \\ 
    \hfil\raisebox{.55em}[0pt]{\bf Potential} & {\bf $\#$DoF }& {\bf Masses} &{\bf $\#$DoF}& {\bf Cosmo}   \\ 
    \hline
    \hline
    $\V[N^A,\g]$ & 6   &\!$m_{0,\ldots,4}^2\neq 0$\! &6=5+ghost$^\dag$\!& $\surd$  \\ \hline  
    $\!\sqrt{\g}\big( N\,\U[\K]+\E[\gamma,\xi]+\U_i\Q^i\big)\!\!$& 5   &  $m_{0}^2= 0$ & 5&$\surd^{\;*}$ \\ \hline  
      As above + Lorentz Invariance\!& 5   &  $m_{0}^2= 0$ & 5& no \\  \hline  
    $\V[N,\g]$ & 3   &$m_{1}^2= 0$&2 & $\surd$  \\ \hline   
    $\sqrt{\g}\,\big(N\,{\bold U}[\g]+{\bold E}[\g]\big)$ & 3   & $m_{0,1}^2= 0$ & 2& $\surd^{\;*}$  \\ \hline  
    $\sqrt{\g}\,\big(\Lambda\,N+{\bold E}[\g]\big)$ & 2   & $m_{0,1,4}^2= 0$ & 2 & $\surd^{\;*}$(CC)\!\! \\ \hline
  \end{tabular}
  \caption{The allowed potentials supporting spatial rotations, 
    and the number of perturbative and nonperturbative  DoF. 
    For perturbative DoF the reference background is Minkowski space. 
    Whether a realistic spatially flat FRW cosmology is admitted is also shown.
    The symbol $^*$ denotes that a further tuning of  the functional
    form of $V$ is required (see condition~(\ref{eq:condcosm})). 
    This tuning is not necessary for time-dependent potentials 
    (see appendix~\ref{app:time} and~\cite{lang}).
    ($^\dag$) the scalar ghost state can become safe on FRW backgrounds 
    (only)~\cite{diego}.}
  \label{tab:summa}
\end{table}

\section{Conclusion}
\label{conc}

We analyzed the Hamiltonian structure of modified gravity theories obtained by adding a
nonderivative function of the ADM variables $\V(N,\,N^i,\,\g_{ij})$ to the Einstein-Hilbert action,
and under the minimal requirement of unbroken rotational invariance, thus encompassing
Lorentz-invariant and Lorentz-breaking theories.  The classification of the various potentials
according to the number of propagating DoF in the perturbative and nonperturbative regime was given
in table~\ref{tab:dof}.  Further restrictions were obtained by requiring the existence of a
realistic FRW cosmology. The results are summarized in table~\ref{tab:summa}.

The simplest deformation, which turns out to propagate \emph{2 DoF}, corresponds to a potential of
the form $\sqrt{g}\,\Lambda +m^2\sqrt{\g}\,{\bf E}[\g]$, i.e.\ a function of the sole 3d metric,
besides the standard cosmological constant. At the level of FRW cosmological background it is
indistinguishable from GR.  Nevertheless, it could lead to possible modifications of gravity in
static solutions. The investigation of the relative phenomenology, for instance of
Schwarzschild-like solutions, is beyond the scope of the present work and will be presented
elsewhere.
 
Potentials that depend on the lapse and the $3d$ metric, $\V[N,\,\g]$, propagate \emph{3 DoF} at
non-perturbative level and they also support FRW solutions where 3 DoF propagate al linear level
(see \cite{diego}).  Unfortunately, only 2 DoF propagate at linearized order around Minkowski
background, indicating strong coupling in the scalar sector.

No potential with \emph{4 DoF} is found. In fact, it seems very difficult if not impossible to
construct $SO(3)$ invariant deforming potential $\V$ with \emph{four DoF}, and so far no nonlinear
realization of partially-massless gravity has been found~\cite{DW1}. Here we showed that, if any
such theory exists, it will appear as a subclass of the Lorentz-breaking potentials with 5 DoF.

The case with \emph{5 DoF} was discussed in depth in~\cite{uscan,uslong,uscosm} and appears to be
promising from a phenomenological point of view, being that the cutoff of the theory is of the order
of $\Lambda_2\sim (mM_{Pl})^{1/2}$ and no vDVZ discontinuity is present.  Although the theory is
weakly coupled with 5 DoF on either Minkowski or FRW backgrounds, the background equations result
incompatible with the requirement of a weakly coupled spectrum on both spaces. Choosing the
existence of FRW spacetime as physical request, one has strong coupling around exact Minkowski and
de Sitter space, with progressively safer cutoff as long as $w_{\text{eff}}$ deviates from $-1$.
Thus, there is a connection between the infrared behaviour of the theory (cosmological scales) with
the short distance behavior (possible short distance strong coupling).  In fact, such a behavior can
set the scale for possible deviations from GR, that may be just around the corner, provided
$w_{\text{eff}}\neq-1$~\cite{uscosm}.  It is important ro remark that Lorentz breaking theories with
5 DoF are immune from issues of spatial (Vainshtein) strong coupling, and thus constitute the first
modified gravity theories for which a weak field expansion is possible.

As is known, the most general potentials, which propagate \emph{6 DoF}, contain the Boulware-Deser
ghost around Minkowski background. Nevertheless, they can support 6 healthy states around FRW
backgrounds~\cite{diego} (see section~\ref{sec:r=4} and~\ref{pert}) and we intend to analyze in
detail the viability of this scenario in a forthcoming work. 

We further found technically possible cases with \emph{5+$\frac{1}{2}$ DoF}, but whether or not one
has to \emph{add or subtract} a half DoF, and under which conditions this has to be considered, is
still an open question \cite{henneaux}. We leave these cases for further investigation.

Finally, our results can be extended to the case of explicitly time-dependent potentials, as
realized if for instance the reference metric is explicitly time dependent. In this case many issues
disappear or are less dangerous (see appendix \ref{app:time}): mainly, the strong coupling around de
Sitter of the models with 5 DoF disappears~\cite{lang}, or, in the case of 2\,DoF, the tuning of the
potentials required to support a FRW background is no longer needed.

\begin{appendix}

\section{Less than Five DoF}
\label{app:4dof}
We  briefly review first (see appendix A in \cite{uslong}) the analysis of the 5 DoF potentials
working with the simpler canonical variables $N,\,\xi^i,\,\g^{ij}$ where the transformation from
$N^i\to \xi^i$ is given by
\be
N^i = N\, \xi^i+\Q^i[\xi,\,\g]\quad {\rm with}\quad
\Q^i=-\left(\partial^2_{\xi^i\xi^j}\U \right)^{-1}\;
\partial_{\xi^j}\E \, .
\ee
As described in the text the potential is of the form
\be\label{5v}
\V=m^2\sqrt{\g}\,\Big(N\,\U+\partial_{\xi^i}\U\,\Q^i+\E\Big) \, ,
\ee
and implies the relations $\partial_N\V=m^2\sqrt{\g}\,\U$,
$\partial_{N^i}\V=m^2\sqrt{\g}\,\partial_{\xi^i}\U$.

The total Hamiltonian in the new variables is
\be
H_T=\int\! d^3x\,\Big[ {\cal H}_A\,N^A+\V+\l^A\;\Pi_A\Big]
=\int\! d^3x\, \Big[({\cal H}_0+{\cal H}_i\,\xi^i)\,N+{\cal H}_i\,\Q^i+\V+\l^A\,\Pi_A\Big] \, , 
\ee
where the momenta $\Pi_0$ and $\Pi_i$, relative to the variables $N$ and $\xi^i$, are now the
primary constraints.
The secondary and tertiary conditions are given by
\bea
\label{eq:Sbari}
{\cal S}_0 &=& ({\cal H}_0+{\cal H}_i\xi^i)+m^2\sqrt{\g}\,\U,\qquad
{\cal S}_i=(N\,\delta^j_i+\Q^j_i)\; \bar \S_j \, ,\\[1ex]
{\cal T}(\vec x)\equiv{\cal T}_0&=&\{\S_0(\vec x),H\},\qquad\qquad\qquad
{\cal
  T}_i(\vec x)=\{\bar\S_i(\vec x),\,H\}+m^2\,\sqrt{\g}\,\U_{ij}(\vec x)\lambda^j(\vec x)
\qquad \\[1ex]
\Rightarrow &&
\l^0 \nb \text{ (undetermined),}\quad\qquad \l^i \text{ (determined)} \, ,
\eea
with $\bar\S_j=({\cal H}_j+m^2\sqrt{\g}\;\U_j)$, which is assumed to vanish to enforce
the constraint $\S_i\approx0$.\footnote{To be rigorous, there exists also a branch of solutions
  to~(\ref{eq:Sbari}) corresponding to $(N\,\delta^j_i+\Q^j_i)\approx 0$. For a simple solvable case
  where $Q^i\equiv \zeta[\gamma]\,\xi^i$ (see~\cite{uslong}) this branches gives 6 DoF and is thus
  unviable, while 5 DoF are present in the branch $\bar \S_i\approx 0$, that we chose.}

The quaternary condition is then
\be
\Q(\vec x)=\{\{\S_0(\vec x),\,H\},\,H\}+\int d^3y\;\left(\l^0(\vec y)\;\{\S_0(\vec x),\,\S_0(\vec y)\}+\l^i(\vec y)\,\partial_{\xi^i(\vec y)}{\cal T}(\vec x) \right)
\ee
and the last lagrange multiplier $\l^0$ is again not determined, completing the elimination of the
sixth mode, provided that 
\be
\{\S_0(\vec x),\,\S_0(\vec y)\}=0 \,.
\ee
This leads to a simple partial differential equation for the potential $\U$, which is solved~\cite{uslong} by the
requirement that $\U$ is a function of the combination $\K^{ij}=\g^{ij}-\xi^i\xi^j$.
Using the expressions for the secondary constraints we can write the Hamiltonian as
\be
H=\int\! d^3x \,\Big[\S_0 \;N+\bar\S_i\,\Q^i+m^2\,\sqrt{\g}\;\E\Big] \, ,
\ee
and we note incidentally that if $\E\equiv0$ then $\Q^i\equiv0$ and $H=\int d^3x\;\S_0\;N$. Thus if
$\{\S_0,\,\S_0\}=0$ then also the tertiary constraint is actually identically zero: ${\cal T} =
\int\!  d^3y\,N(\vec y)\,\{S_0(\vec x) ,\,S_0(\vec y) \}\equiv 0$.  In this case the Dirac analysis
stops at the level of secondary constraints, and one is left with 6 DoF instead of 5.  Therefore,
$\E\neq 0$ is a necessary conditions to have 5 DoF.

\medskip

We can now find the further conditions under which a potential of the form (\ref{5v})\ propagates
only 4 DoF.  One must require that even the quinary and the senary conditions 
\bea
 \F=\{\Q,\,H\}+\l^i\;\partial_{\xi^i}\Q+\l^0\;\partial_N\Q \,,\\
 {\cal G}=\{\F,\,H\}+\l^i\;\partial_{\xi^i}\F+\l^0\;\partial_N\F 
\eea
do not determine the last lagrange multiplier $\l^0$.  This implies the following partial
differential conditions in field space
\bea
 \partial_N \Q=0\,,\qquad {\rm and} \qquad \partial_N  {\F}=0 \,.
\eea 
The explicit expressions consist in rather complicated equations for $\U$ and $\E$; no solution,
if any exists, is presently known.

The last Lagrange multiplier ought to be determined at the next (septenary) step. However as usual,
even less DoF may be present. For instance, gauge invariance could be present, if the lagrange
multiplier is not determined even by the septenary and octonary conditions and the procedure stops
there.  The number of DoF would in this case be 3, with both second and first class constraints.

\section{Explicitly time dependent potentials}
\label{app:time}
Here we analyze the case where the potential $\V$ has an explicit time dependence allowed by spatial
rotational $SO(3)$ symmetry (see \cite{lang} where the class of potentials with 5 DoF was
considered).  For what concerns the existence of a FRW background, the Bianchi conservation equation
(\ref{bianchi}) acquires an extra term, namely
\be
\label{bianchi1}
\dot N\;\V_{NN}- 
6\,{\frac{\dot a}{a}}
\left(\V_{N\g}-\frac{\V_\g}{N}\right)+N\,\partial_t\V_N=0 \, .
\ee

For the potentials of the form $\V=\sqrt{\g}\,(N\hat{\cal A}+\hat{\cal B})$, with $\hat{\cal A}$ and
$\hat{\cal B}$ explicitly time dependent, equation (\ref{bianchi1}) becomes an algebraic equation for
the lapse $N$ (see \cite{lang})
\be
\label{eqN}
6\,{\frac{\dot a}{a}}\left( \hat {\cal B}_\g-\frac{1}{2}{\hat{\cal B}}
\right)+N\,\partial_t{\hat{\cal A}}=0 \, .
\ee
The last term can be understood by the fact that the explicit time dependence of $\hat{\cal A}$ is also a
source of breaking of time reparametrization.  In any case, now $N$ will be determined, and from
the 00 component of the Einstein equations, $3(\frac{\dot a}{a})^2=H^2=N^2\rho_{\text{eff}}$, one
can determine the scale factor $a$, leading to a sensible cosmology.

An straightforward construction leading to explicit time dependence is provided by a nontrivial
spatial reference metric, if one replaces $\delta_{ij}$ with $b(t)\delta_{ij}$,
see \cite{lang}.  All invariants are built from the spatial tensor $\g^{ik}\,\delta_{kj}$ and the
net effect is to replace $\g^{ij}$ by $b(t)\,\g^{ij}\equiv\tilde\g^{ij}$.  In this case, we can
write eq.(\ref{eqN}) as: 
\be
N= 2\,\frac{\dot a/a}{\dot b/b} \,\frac{\hat{\cal B}_\g-\hat{\cal B}/2}{\hat{\cal A}_\g}\,,
\ee
where $\partial_t\hat{\cal A}=\partial_t\tilde\g^{ij} \,\partial_{\tilde\g^{ij}}{\hat{\cal A}}$ and
$\partial_{\tilde\g^{ij}}{\hat{\cal A}}\equiv {\hat{\cal A}}_\g\tilde\g_{ij}$ and
$\partial_t\tilde\g^{ij} =\dot b\,\g^{ij}$.  Using comoving time $\tau$ (i.e. $N\,d t=d\tau$)
equation (\ref{eqN}) remains the same (and determines $N(\tau)$) while the 00 Einstein eq. becomes
$3\frac{a'(\tau)^2}{a(\tau)^2}=8 \pi G \, \rho(\tau) + \rho_{\text{eff}}(\tau)$. Note that Minkowski
space is not solution of the modified Einstein equations. Also, note that the limit of static $b$ is
singular, and one turns back to the constraint ${\cal B}_\g-{\cal B}/2=0$ as in the text.

\medskip

The analysis reported in the present work, extended to time dependent potentials, gives basically the
same results summarized in table~\ref{tab:summa}, but, despite the need to introduce an arbitrary
time-dependent function, it represents an interesting possibility, because it avoids the tuning
condition for the functional form of $\V$ required to have a FRW background.  Moreover, for class of
potentials with 5 DoF, the theory is weakly coupled also near de Sitter backgrounds~\cite{lang} and
thus represent a viable massive gravity with 5 DoF.

\medskip

We finally carried out a check that all the nonperturbative results mentioned in this work can be
extended to the case of an explicit time dependence, by repeating the analysis of constraints for
the case of 5 DoF.  It is straightforward to check that primary ($\Pi_A$) and secondary constraints
(${\cal S}_A$) are not affected by the explicit time dependence. This is the main reason why our
results can be safely extended.  Tertiary constraints are modified according to
\be
\hat{\cal T}= {\cal T}+m^2\,\partial_t \big (\sqrt{\g}\,\U \big),\qquad \hat{\cal
  T}_i={\cal T}_i+\partial_t\,
\bar{\cal S}_i \, ;
\ee
where we have denoted with $~\hat~~$ the corresponding constraints in the case of explicit time
dependence, see appendix~\ref{app:4dof}. The quaternary constraint becomes
\be
\hat{\cal Q}={\cal Q}+m^2\,\big\{ \partial_t \big (\sqrt{\g}\,\U \big),\,H \big\}+m^2\,\partial_t^2 \big(\sqrt{\g}\,\U \big) \, .
\ee
The lagrange multiplier $\l^0$ is again not determined at this stage if $\{{\cal S}_0,\,{\cal
  S}_0\}=0$, which is the same partial differential equation for the potential $\U$ as in the
time-independent case. Thus, we again find that the potentials of the form
\be
\V\equiv m^2\,\sqrt{\g}\,\Big( N\;\hat \U +\partial_{\xi}\hat\U\;\hat {\cal Q}^i+\hat\E \Big)\,,
\ee 
with $\hat\U=\U[{\cal K}^{ij},t]$ and $\hat\E=\E[\g^{ij},t]$, propagate nonperturbatively 5 DoF.
For the cases with less than 5 DoF, all the results can be extended along the same lines.

\end{appendix}

 \section*{Acknowledgements}

L.P. thanks the Cosmology and Astroparticle Physics Group of the {\it University of Geneva} for
hospitality and support.  D.C. dedicates the present work to: "{\it Luigi delle Bicocche, eroe
  contemporaneo a cui noi tutti dobbiamo la nostra libert\`a\ldots  ({\it Eroe}, Caparezza).}" having
inspired part of his job.

%

\end{document}